\documentstyle[aps,twocolumn,tighten]{revtex}

\input epsf
\def\plotone#1{\centering \leavevmode
\epsfxsize= 1.0\columnwidth \epsfbox{#1}}

\def\be{\begin{equation}}
\def\ee{\end{equation}}
\def\bea{\begin{eqnarray}}
\def\eea{\end{eqnarray}}

\def\cmm2{{\,\rm cm^{-2}}}
\def\cm2{{\,{\rm cm}^2}}
\def\cmm3{{\,{\rm cm}^{-3}}}
\def\gcmm3{{\,{\rm g\,cm^{-3}}}}

\def\fun#1#2{\lower3.6pt\vbox{\baselineskip0pt\lineskip.9pt
\ialign{$\mathsurround=0pt#1\hfil##\hfil$\crcr#2\crcr\sim\crcr}}}

\def\eg{{e.g., }}
\def\ie{{i.e., }}

\def\p3m{P$^3$M}

\def\la{\mathrel{\mathpalette\fun <}}

\def\fun#1#2{\lower3.6pt\vbox{\baselineskip0pt\lineskip.9pt
  \ialign{$\mathsurround=0pt#1\hfil##\hfil$\crcr#2\crcr\sim\crcr}}}

\begin{document}
\twocolumn[\hsize\textwidth\columnwidth\hsize\csname @twocolumnfalse\endcsname
\draft
\title{CMB Power Spectrum Estimation via Hierarchical Decomposition} 
\author{Olivier\ Dor\'e$^{1}$, Lloyd\ Knox$^{2}$, 
and Alan\ Peel$^{2}$}
\address{$^1$ Institut d'Astrophysique de Paris,
98bis Boulevard Arago, F-75014 Paris, FRANCE; dore@iap.fr}
\address{$^2$ Department of Physics, One Shields Avenue,
University of California, Davis, California 95616, USA;
\\ lknox@ucdavis.edu, apeel@bubba.ucdavis.edu}
\date{\today}
\maketitle

\begin{abstract}

We have developed a fast, accurate and generally applicable method for
inferring the power spectrum and its uncertainties from maps of the cosmic
microwave background (CMB) in the presence of inhomogeneous and correlated
noise.  For maps with $10^4$ to $10^5$ pixels, we apply an exact power
spectrum estimation algorithm to submaps of the data at various
resolutions, and then combine the results in an optimal manner.  To
analyze larger maps efficiently one must resort to sub--optimal
combinations in which cross--map power spectrum error correlations are
only calculated approximately.  We expect such approximations to work well
in general, and in particular for the megapixel maps to come from the next
generation of satellite missions.

\end{abstract}
] 
\section{Introduction}

The anisotropy of the Cosmic Microwave Background (CMB) is proving to be a
powerful cosmological probe \cite{jaffe00}.  Many cosmological parameters,
and the primordial power spectra of density and gravity--wave
perturbations, can be inferred from the statistical properties of the
CMB---in particular from its angular power spectrum \cite{forecast}.  
Unfortunately, exact methods for calculating the power spectrum and its
uncertainties from real observations are very expensive
computationally\cite{bcjk}.  Supercomputers are required for analysis of
current datasets and even they will not be sufficient for the next
generation of experiments \cite{borrill99}.  Here we introduce an
approximate method for reducing a CMB map to a power spectrum and its
uncertainties.

Generally applicable exact methods for finding the angular power spectrum,
$C_l$, that maximize the likelihood have operation counts proportional to
$N^3$ where $N$ is the number of pixels in the map.  Our approach to
overcoming this $N^3$ scaling involves a hierarchical decomposition of the
map into a set of submaps. That is, we subdivide the original
(``primary'') map into non--overlapping regions, each with a manageable
number of pixels, in order to estimate the power spectrum from each of
these submaps using an exact algorithm.  To study the larger angular scale
fluctuations we coarsen the primary map and if the number of these coarse
pixels is still too large, we again divide into submaps.  To go to yet
larger angular scales, we coarsen the map further, etc...  Then we
calculate the expected correlations between the power spectrum estimates
from all different submaps at all different resolution levels in order to
optimally average them together.  A similar multi--grid technique was
recently developed for the reduction of time--ordered CMB data to
maps\cite{dore01}.

Several other approaches to overcoming the $N^3$ scaling have been tried.  
These include the ``pseudo--$C_l$'' method of \cite{wandelt}, and the
``correlation--function'' approach of \cite{szapudi00}.  We expect these
methods to work well in the case of homogeneous noise, but to be
significantly sub--optimal for the levels of inhomogeneity expected in
planned observational programs.  None of these methods has been shown to
deal properly with correlated noise.  Minor modifications of the
correlation--function approach may make this path very attractive, though
a remaining issue is the importance of noise correlations between pixels.

The $N^3$ scaling has been overcome also by a special--purpose exact
method that is expected to be applicable to the maps generated by NASA's
Microwave Anisotropy Probe ({\it MAP\/})
satellite\footnote{\texttt{http://map.gsfc.nasa.gov/}}.  This method
\cite{oh99} assumes the noise is not correlated from one pixel to another
and that the noise level variations are roughly azimuthally symmetric.  
Some of the techniques used in \cite{oh99} may eventually find their place
in more generally applicable (and yet still exact) power spectrum
estimation algorithms, though the feasibility is not yet clear.  Another
special--purpose exact algorithm is that of \cite{wandelt01}, which is
applicable to experiments that scan on rings.  The main idea is to analyze
ring sets instead of maps since both the noise and signal covariance
structures are simple on the rings, whereas the noise structure can be
complicated in the map space.  Although some of its critical hypotheses
have not been tested yet on realistic data, the ring--set approach might
still be of practical importance since it may provide a useful
zeroth--order solution for experiments that nearly scan on rings.

In section II we describe our method in detail.  In section III we present
the results of an application to a map with ten thousand pixels
---comparable to the size of maps coming from long--duration balloon (LDB)
flights.  In section IV we show results from a map four times larger and
discuss prospects for application of our method to even larger maps such
as those expected from {\it MAP\/} and {\it
Planck\/}\footnote{\texttt{http://astro.estec.esa.nl/SA-general/Projects/Planck/}}.  
In section V we compare with other methods. In section VI we conclude.

\section{Method} 

Here we first describe our method in the simplest conceptual terms, and
then go on to discuss subtleties which complicate our implementation.

\subsection{From the Likelihood Function to the Quadratic Estimator}

We describe here the use of a quadratic estimator to find the maximum of
the likelihood function, and the shape of the likelihood function near
that maximum, as described in \cite{bjk98}.  Time--ordered data from
observation of the CMB are usually reduced to a set of pixelized maps
$\Delta_i$, $i=1,\ldots N$ which can be decomposed into the sum of a
signal and a noise contribution, $ \Delta = s + n \ $.  Assuming that both
the noise and the signal are normally--distributed, their statistical
properties are fully characterized by the covariance matrices $S = \langle
s s^T \rangle$ and $N = \langle n n^T\rangle$.  Assuming furthermore that
the noise and signal are not correlated with each other, we can define
\be
C \equiv \langle\Delta \Delta^T\rangle = S + N \; .
\ee

The observed sky signal is assumed to be the realization of an isotropic
Gaussian random field whose power spectrum $C_l$ is the quantity we
want to measure.  Thus we are interested in the likelihood function
${\mathcal L}(\Delta \mid C_l)$ which is given by
\be
-2 \ln {\mathcal L}(\Delta \mid C_l) = \ln\
\mathrm{det}\ C + \Delta^TC^{-1}\Delta \; .
\ee
In particular we are interested in the location of the maximum of this
function (which is the most likely $C_l$) and the curvature at the
maximum, $-\partial^2\ln{\mathcal L}/\partial C_l \partial C_{l'}$ (which
is approximately the inverse of the covariance matrix for $C_l$). Note
that $C$ depends on $C_l$ since
\be
S_{ii'} = \sum_{\l} {2\l +1 \over 4\pi} C_l {\mathcal W}_{ii'}(l)
\ee
where $\mathcal{W}$ is the covariance window function of the experiment.  

Given an initial estimate of $C_l$ (hereafter, the input $C_l$) one can
reach the likelihood maximum as follows.  By Taylor--expanding $\ln
{\mathcal L}$ to second order in $\delta C_l$ around $C_l$, and replacing
$-\partial^2\ln{\mathcal L}/\partial C_l \partial C_{l'}$ with its
expectation value one can find an expression for $\delta C_l$ such that
$C_l + \delta C_l$ maximizes the likelihood:
\be
\label{eqn:quadest}
\delta C_l = \sum_{l'} {1\over 2}F^{-1}_{ll'}{\rm Tr}\left[\left(\Delta
\Delta^{\rm T} - C\right)\left(C^{-1}{\partial C \over \partial C_{l'}}
C^{-1}\right)\right]
\ee
and 
\be
\label{eqn:fish}
F_{ll'} 
= {1\over 2}{\rm Tr}\left[C^{-1}
{\partial C \over \partial C_l}
C^{-1}{\partial C \over \partial C_{l'}}\right]
\ee
is the Fisher matrix \cite{tegmark97}.  

Equation~\ref{eqn:quadest} is a quadratic function of the data and hence
the expression ``quadratic estimator''. Note that we have suppressed the
pixel indices in the various vectors and matrices.  Since $\ln{\mathcal
L}$ is not equal to its second--order Taylor expansion (i.e., ${\cal L}$
is not a Gaussian in $C_l$), some iteration is generally required to reach
the likelihood maximum.

\subsection{Hierarchical Decomposition and Recombination}

Now let us consider multiple maps and use Greek indices to label them.
Estimates of $\delta C_l$ from map $\alpha$, $\delta C^\alpha_l$, are
correlated with those from map $\beta$ with correlation matrix:
\bea
\label{eqn:correl}
\langle \delta C^\alpha_l \delta C^\beta_{l'} \rangle & \equiv &
{\mathcal F}^{-1}_{\alpha l,\beta l'} \\
& = & \sum_{l'',l'''} (F_\alpha^{-1})_{ll''}(F_\beta^{-1})_{l'l'''}
\times \nonumber \\
& & \ \ \ {1\over2}{\rm Tr}
\left[A_{\alpha,l''}C_{\alpha\beta}A_{\beta,l'''}C_{\beta\alpha}\right]\nonumber
\eea
where
\be
\label{eqn:defA}
A_{\alpha,l} \equiv C_{\alpha\alpha}^{-1}{\partial C_{\alpha\alpha} \over 
\partial C_l}C_{\alpha\alpha}^{-1}\; .
\ee
Note that if $\alpha =\beta$ then Eq.~\ref{eqn:correl} simplifies to the
usual result:
\be
\langle \delta C_l \delta C_{l'} \rangle = F_{ll'}^{-1}\; .
\ee

Given this result, we know how to combine the various $\delta C_l$
estimates from each submap into a final $\delta C_l$ estimate from all the
submaps in a minimum--variance (optimal) manner.  The minimum--variance
$\delta C_{l'}$ satisfies
\be
\label{eqn:combine}
\sum_{l'}\left(\ \sum_{\alpha \beta}\ {\mathcal F}_{\alpha l,\beta l'}\
\right)\ \delta C_{l'}\ =\ \sum_{\alpha \beta l'}\ {\mathcal F}_{\alpha
l,\beta l'}\ \delta C^\alpha_{l'}
\ee
and has a weight matrix (inverse of covariance matrix):
\be
F_{ll'} = \sum_{\alpha \beta} {\mathcal F}_{\alpha l,\beta l'}\; .
\ee

Although for simplicity we have written these expressions for estimating
individual $C_l$'s, issues of signal--to--noise and spectral resolution
usually lead us to estimate the power spectrum in bands of $\ell$, where
the shape of $C_l$ inside the bands is assumed.  The usual assumption
(which we use in our applications) is that $l(l+1) C_l/(2\pi)$ is constant
inside the band.

Our treatment of the correlations of the $\delta C_l$'s between pairs of
maps has been general.  The maps may be spatially separate or overlapping;
they may have equivalent or different pixel sizes.  Thus we have worked
out the most general solution to optimally combine the power spectra of
submaps which are the result of hierarchical decomposition (HD) of a
primary map.

\subsection{Spectral Resolution}

Even with optimal combining of the power spectrum estimates from the
various submaps, the HD procedure results in a sub--optimal estimation of
the power spectrum.  Fortunately, in the cases we study, the sub--optimal
results are quite close to the optimal results.  Departure from the
optimal results is almost entirely due to the degraded spectral resolution
of the high--resolution submaps. This loss of spectral resolution is the
primary drawback of the HD approach.

The spectral resolution is most severely degraded at the highest
resolution levels where the submaps have the smallest spatial extents.  
For any map of linear extent, $L$, it is difficult to distinguish the mode
$P_l(\cos\theta)$ from $P_{l+\delta l}(\cos\theta)$ where $\delta l \la
\pi/L$ \cite{endnote5}.  If one wishes to achieve a spectral resolution of
$\delta l$ for a square map with linear pixel size $r_p$ then this map
must have $n$ pixels where
\be
n \simeq 2.5 \times 10^3 \left( {30 \over \delta l} {7' \over r_p}
\right)^2\; .
\ee
Fortunately $\delta l = 30$ and $r_p = 7'$ are expected to be adequate for
LDB--type maps and $2.6 \times 10^3$ pixels is a small enough submap size
to allow for reasonable computation times (as shown below).

\subsection{Scaling}

We now calculate how computation time scales with total number of pixels
in the full--resolution primary map, $N$, and the number of
multipole--moment bands, $N_b$.  For simplicity we assume that all submaps
have the same number of pixels, $n$, and that we estimate the power
spectrum for each submap in the same number of bands. Estimating the power
spectrum and Fisher matrix for each submap takes on the order of
$N_b^2n^3$ operations so at the finest level we have on the order of
$N_b^2(N/n) n^3 = N_b^2N n^2$ operations.  In a systematic coarsening
(such as the one below defined by combining four pixels at one resolution
into one larger pixel for the next coarser level), most of the submaps are
at the finest resolution and therefore analysis and combining of these
finest submaps dominates the demands on memory and CPU time.

For large enough $N$, the dominant computational step will be in
calculating the correlations between submaps.  The matrix multiplication
in Eq.~\ref{eqn:correl} takes on the order of $n^3$ operations, so
performing it for every pair of submaps and pair of bands takes on the
order of $N_b^2N^2 n$ operations.

The procedure can in principle be parallelized for the efficient use of
$n_{\rm proc}$ processors, where $n_{\rm proc}$ ranges anywhere from $N_b$
to $\sim (N/n)^2 N_b^2$.  The crucial use of parallelization comes in the
dominant combining stage, which scales as $(N/n)^2 N_b^2$, and involves
the combination of $\sim {1\over 2}(N/n)^2$ pairs of submaps.  This type
of independent pair loading can be efficiently shared on any number of
processors lower than ${1\over 2}(N/n)^2$. For LDB--type missions one
might have $N_b \sim 10$ and $(N/n)^2 N_b^2 \sim 4\times 10^4$ and
approximately $200$ pairs of submaps to combine (if we indiscriminately
retain all submap-submap correlations (see Section IV)).  For
supercomputers with $n_{\rm proc} \la 10^2$, every processor can be
efficiently used.

\subsection{The Noise Matrix}

Our approach assumes that we begin with a pixelized map and its
corresponding noise covariance matrix.  Map--making procedures usually
produce a weight matrix, which is the inverse of the noise matrix.  
Inverting an arbitrary weight matrix takes on the order of $N^3$
operations. Fortunately, this inversion only needs to be done once and is
feasible for LDB--size maps.

For larger maps, treatment of the weight matrix by general matrix
inversion algorithms is impossible.  Fast methods are being developed
\cite{borrill01} which rely on the origin of the map weight matrix in the
weight matrix of the time--ordered data.  That is, the map weight matrix
is $A^TN^{-1}A$ where $N^{-1}_{tt'}$ is (here) the time--stream weight
matrix for a stationary noise process, and $A_{ti}$ is the pointing matrix
element that is one if at time--sample $t$ the telescope is sampling map
pixel $i$ and zero otherwise.  This special structure allows for each
iteration of a conjugate gradient solution to be performed much faster
than for an arbitrary matrix.

Another possibility (suggested in \cite{szapudi00}) is to calculate the
noise covariance matrix by Monte Carlo methods.  In other words, one would
make repeated simulations of the map noise and average those together to
get any desired elements of the noise matrix.  In addition to possible
speed advantages, this approach also has storage advantages since one
probably needs fewer than $N/2$ realizations to have a sufficiently
accurate estimate of the noise.  One may still need thousands of
realizations of the noise---\eg 20,000 realizations are required for the
noise matrix elements to be accurate to within 1\% of the diagonal.

\subsection{Coarsening}

The amount of work to be done depends on the choice of number of
resolution levels, which is governed by how many pixels are combined to
form one pixel at the next--coarsest level.  Greater coarsening between
levels leads to fewer required operations, but at the expense of greater
loss of information.  Since the cost in computing time is slight for using
the most modest coarsening possible while maintaining (roughly) square
pixels, we always coarsen by averaging four pixels into one.  This
coarsening is also easily implemented in the HEALPix pixelization scheme,
which we use \cite{GoHi98}.

In general, one can create a coarse submap $\Delta$ from a fine submap
$\delta$ as follows:
\be
\label{eqn:coarsen}
\Delta = W^{-1} \alpha w \delta 
\ee
where $\alpha_{ci}$ is one for all fine pixels $i$ in coarse pixel $c$ and
zero otherwise, $w$ is some weighting of the fine pixels and $W = \alpha w
\alpha^T$.  The coarse--fine and coarse--coarse noise covariance matrices
are given by:
\bea
\label{eqn:finecoarsenoise}
\langle \Delta \delta^T \rangle_{\rm noise} &=& W^{-1}\alpha w N \\
 &=& W^{-1}\alpha \ \ ({\rm if \ }w=N^{-1}) \nonumber \\
 &=& {N\alpha\over 4} \ \ ({\rm if \ }w=I) \nonumber 
\eea
and:
\bea
\label{eqn:coarsecoarsenoise}
\langle \Delta\Delta^T \rangle_{\rm noise} &=&
W^{-1}\alpha wNw^T\alpha^T W^{-1} \\
 &=& W^{-1} \ \ ({\rm if}\ w=N^{-1}) \nonumber \\
 &=& {\alpha N\alpha^T\over 16} \ \ ({\rm if}\ w=I)\; . \nonumber
\eea

For optimal coarsening $w=N^{-1}$ and for uniform averaging, $w=I$.  We
assume that we are coarsening four pixels into one and therefore that
$\alpha \alpha^T = 4I$.  We see that uniform averaging leads to noise
covariance matrices that are easy to calculate.  For optimal averaging we
need to invert $W$ which is substantially less challenging than inverting
$N^{-1}$ to get $N$ since it has 1/16 the number of elements. The same
technique used for calculating $N$ by exploiting the origin of $N^{-1}$ in
time--ordered data (as explained in the previous subsection) can be used
to get $W^{-1}$ \cite{jaffe01}.

Coarsening will usually result in pixel sizes that are large compared to
the angular resolution of the instrument and therefore pixelization
effects must be taken into account.  Our treatment of the effect of
pixelization on the signal correlation function is approximate, \ie we use
a pixel window which is the average of the evaluated power spectrum for
every individual pixel.  To prevent these approximations from creating
errors in the final power spectrum, we ignore information from multipole
moments greater than some critical value where the approximation
introduces significant error.  Pixelization effects are discussed in more
detail in the Application section.

\subsection{Iteration}

A single application of the quadratic estimator of Eq.~\ref{eqn:quadest}
might not result in a $C_l$ that is sufficiently close to the likelihood
maximum.  This will be the case if the input $C_l$ is too far from the
likelihood maximum.  Fortunately, iterative application of
Eq.~\ref{eqn:quadest} has been shown to converge quite rapidly
\cite{bjk98}.

When using the hierarchical decomposition approach, it is important that
the iteration be done {\it globally}.  That is, within each iteration, the
power spectrum from each submap should be estimated using the same input
$C_l$.  If iteration is performed within the submaps, the combined result
will suffer from cosmic bias\cite{bjk00}, which results from the fact that
uncertainties in $C_l$ are not normally--distributed.  For a
normally--distributed variable, the curvature of the log of the likelihood
function is independent of location in the parameter space (because the
likelihood is a Gaussian).  However, for $C_l$, this curvature does depend
on location.  For larger values of $C_l$ the curvature is smaller (\ie the
variance is larger).  Thus, upward fluctuations should result in larger
variances than downward fluctuations and so if one combines them together
assuming Gaussianity, the net result is a downward bias due to the
over--weighting of the downward fluctuations.

The combination procedure of Eq.~\ref{eqn:combine} implicitly assumes the
estimates are normally distributed.  We avoid the cosmic bias that might
result from this assumption by weighting the downward and upward
fluctuations equally. That is, we make sure to calculate ${\mathcal
F}_{\alpha l,\beta l'}$ from the same $C_l$ for all submaps.  Thus any
desired iteration, \eg motivated by a large correction from the input
$C_l$, should be done globally.

Since the uncertainty in $C_l$ is non--Gaussian, specifying the $C_l$ that
maximizes the likelihood function, and $\langle \delta C_l \delta C_{l'}
\rangle$, does not completely characterize the uncertainty.  The
uncertainty can be approximately characterized by use of the ``offset
log--normal form'' \cite{bjk00}.  That is, error in the quantity $Z_l
\equiv \ln\left(C_l + x_l\right)$ {\it is} approximately
normally--distributed.  The offset, $x_l$, is a measure of the noise
contribution to the uncertainty, as opposed to the sample--variance
contribution to the uncertainty.  It can be calculated as outlined in
\cite{bjk00}.

\section{Application}

First we discuss the specifications for the simulated maps we used. Then
we compare the results of HD with those of the exact method.

\subsection{Simulation Map Details}

We have applied our method using a Fortran code, which we have named {\it
Madcumba}, to two different simulated maps, hereafter simulations $A$ and
$B$. In both cases, the angular--power spectrum used was that of a {\it
COBE}--normalized adiabatic, scale--invariant ``lambda'' cold dark matter
($\Lambda$CDM) model with $\Omega_\Lambda = 0.6$, $\Omega_b = 0.05$,
$\Omega_{\rm cdm}=0.35$ and $H_0 = 75\ {\rm km\,sec^{-1}\,Mpc^{-1}}$ and
was generated by the publicly available code CMBfast \cite{seljak96}.  
The simulated signal maps were generated using the \texttt{synfast}
routine in the publicly available HEALPix package \cite{GoHi98}, at
HEALPix $N_{\rm side} = 256$ (level 8, where $N_{\rm side} = 2^{\rm
level}$), in which the pixel solid angle is around $(13.7')^2$, assuming a
circular beam with full--width at half--maximum of $20'$.  Finally, pixel
noise taken from a Gaussian distribution with zero correlations between
pixels was added to the maps. The only significant differences between our
two simulations are size and noise characteristics.

The simulation $A$ map has $10^4$ pixels, is square in shape, and has a
homogeneous noise variance of $(20\ \mu K)^2$ in each pixel.  Its
relatively small size allows for the power spectrum to be estimated by the
exact method (\ie without dividing into submaps) using the MADCAP
package\cite{borrill99}.  This is compared to our calculation via HD into
four equal--area square 2500 pixel submaps at full resolution and one
coarse 2500 pixel submap at HEALPix $N_{\rm side} = 128$ (level 7) which
covers the same area as the primary map.

The simulation $B$ map is also square in shape and has $4\times 10^4$
pixels with a noise variance that is cosine--modulated throughout the map,
varying from $(20\ \mu K)^2$ to $9 \times (20\ \mu K)^2$. Here, we
decompose the primary map into sixteen submaps at full resolution, four
submaps at the next coarser resolution and one coarsest resolution submap
which covers the same area as the primary map but, by being two levels
coarser, contains $1/16^{\rm th}$ as many pixels. Thus, as with simulation
$A$, we use $n=2500$ pixel submaps.

\subsection{Comparison with Exact Method}

\begin{figure}
\plotone{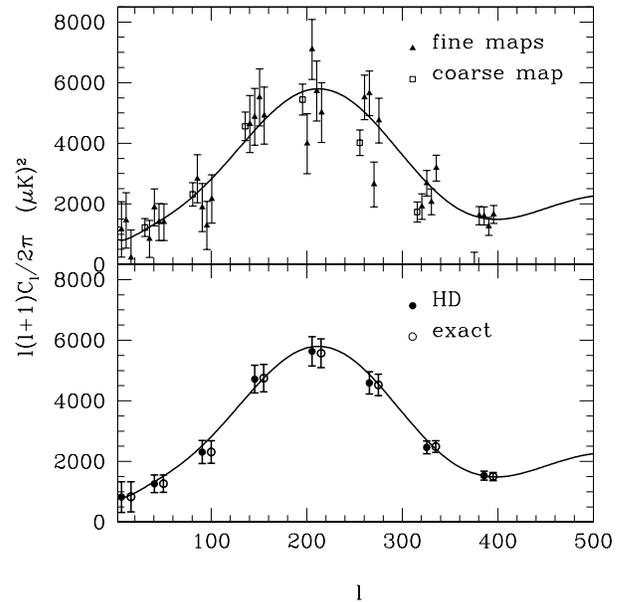} \caption{Simulation $A$ Results.  Top panel:  Power
spectrum estimates from four individual full--resolution 2500 pixel
submaps (triangles) and one coarse 2500 pixel submap.  Bottom panel:  
Power spectrum estimates from optimally combining the top--panel results
(solid circles) and from the exact calculation (open circles).  Note that
in both panels, points are slightly shifted horizontally for
clarity.\label{fig:pow_smallmap}} 
\end{figure}

The top panel of Fig.~\ref{fig:pow_smallmap} shows estimates of the powers
from the individual submaps in simulation $A$.  The bottom panel shows
both the result of optimally combining them and the exact results obtained
directly from the primary map.  The solid line in both panels is the
original power spectrum for the simulations.  The differences between the
power estimates are less than 20\% of the standard error from the exact
method.

\begin{figure}
\plotone{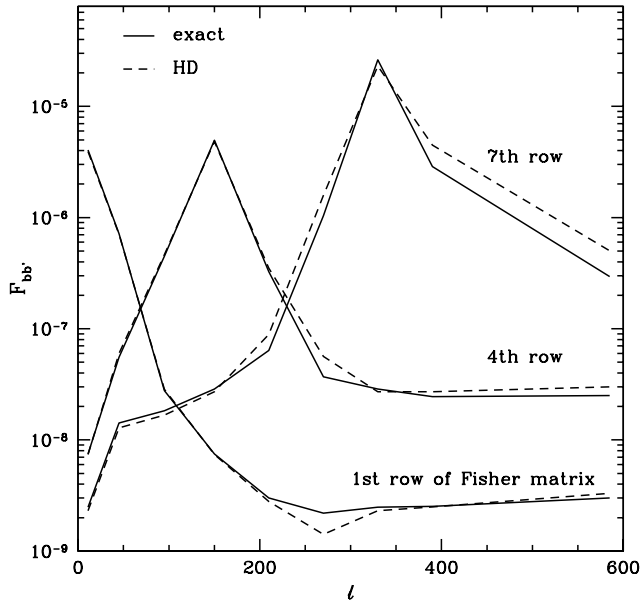}
\caption{\label{fig:fish_smallmap}Three rows of the Fisher matrix
calculated exactly (solid lines) and also via the combination (HD)
procedure (dashed lines) for simulation $A$.}
\end{figure}

Not only do the power spectrum estimates agree quite well, but so do the
estimates of the uncertainties.  The error bars in
Fig.~\ref{fig:pow_smallmap} are the square roots of the diagonal elements
of the respective Fisher matrices.  In Fig.~\ref{fig:fish_smallmap} one
can see how well entire rows of the exact and HD Fisher matrices agree.

\begin{figure}
\plotone{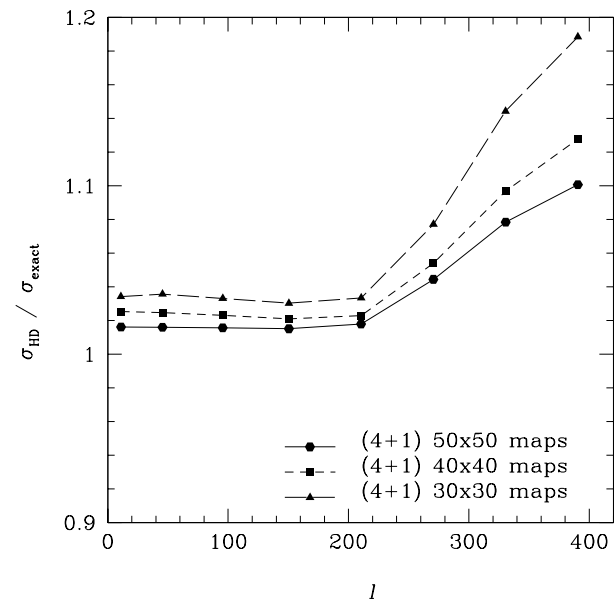}
\caption{Error bars from HD divided by error bars for the exact analysis.
Each case represents a primary map with 4$n$ pixels divided into four
$n$--pixel full resolution submaps and one coarsened $n$-pixel map where
$n = 30 \times 30$ (triangles), $40 \times 40$ (squares) or $50 \times 50$
(hexagons).
\label{fig:sigmab}}
\end{figure}

Clearly, the bigger the submaps at the finest resolution, the better this
approach works.  For a fixed length scale of interest, larger submaps
contain a greater fraction of corresponding pixel pairs, and therefore
achieve better spectral resolution ($\delta \ell$).  Unfortunately, the
compute--time, when dominated by the combine procedure, scales as $n$ and
therefore as $1/\delta \ell^2$ (or possibly $n^2$ but with a much smaller
pre--factor (see section IV)). Thus, choice of $n$ can be critical.  We
studied how our information loss varies with $n$ by comparing the error
bars from the HD procedure to the full analysis for $n=2500$ (the case
above), $n=1600$, and $n=900$.  The results are shown in
Fig.~\ref{fig:sigmab}.  Note that for the $n=2500$ case all the error bars
are increased over the exact case by less than 10\%.  These larger error
bars are consistent with the less than 20\% differences (in units of
variance of exact results) between the power estimates. 

\begin{figure}
\plotone{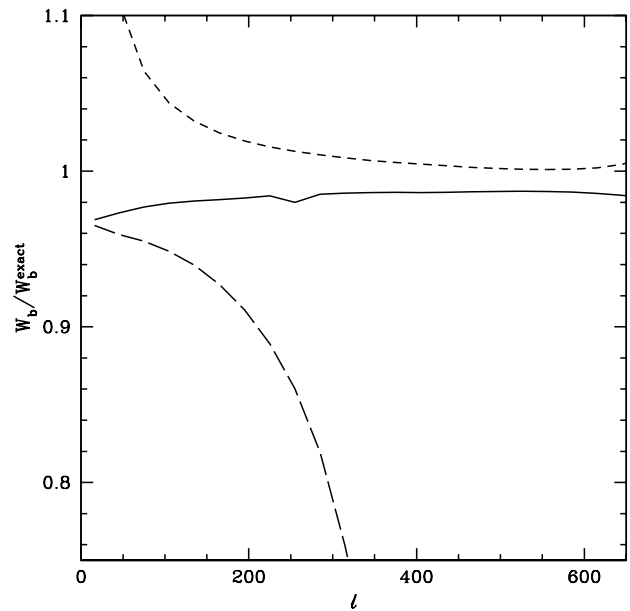}
\caption{$W_b/W_b^{\rm exact}$ where $W_b = \sum_{b'}F_{bb'}$. The $W_b$'s
are from analysis of the simulation $A$ map, but into finer bins of width
$\delta l = 30$.  The short--dashed line is $4W_b/W_b^{\rm exact}$ where
$W_b$ is just from analysis of one of the four full--resolution submaps;
the long--dashed line is $W_b/W_b^{\rm exact}$ where $W_b$ is from
analysis of the coarse resolution submap; the solid line is $W_b/W_b^{\rm
exact}$ where $W_b$ is from combining information from all five submaps.
\label{fig:weight}}
\end{figure}

The upward trend in error ratio with increasing band number is an effect
of decreasing spectral resolution.  To understand this, we examine
Fig.~\ref{fig:weight} which shows the ratio of the HD over the exact
method of the band contributions to the total weight, $W_b$, where:
\be
W_b \equiv \sum_{b'} F_{bb'}
\ee
and the total weight of an experiment is $W \equiv \sum_{b} W_b$. For this
analysis we switch to a finer binning of 25 bands, each with width $\delta
l = 30$.

Note first the short--dashed line which is four times the ratio of $W_b$
for one full resolution submap over the one for the primary map.  If the
four submaps were uncorrelated, we would expect this ratio to be $\sim 1$.  
However, since the submaps are correlated, this ratio is greater than 1.
We see that submap--submap correlations are more important at lower $\ell$
than higher $\ell$ values.

Though individual elements of the Fisher matrix may be larger for a
sub--optimal method than an optimal one, we know that the contribution
from a given band to the total weight {\it can not} be larger.  Thus, the
best we could hope for is that the ratio of $W_b$ for the HD method over
the exact method is near unity.  We see from Fig.~\ref{fig:weight} that it
is everywhere greater than 0.97.  Thus the fact that the combine procedure
gives at most 10\% larger error bars (20\% larger variances) in
Fig.~\ref{fig:sigmab} can not be due to any reduction in the total weight
(which we see is negligible), but must be due to how each $W_b$ is
distributed among the $F_{bb'}$.  In particular, it is the lower spectral
resolution of the smaller submaps which results in the $W_b$ being more
spread out within a Fisher matrix row and less concentrated in the
diagonal element $F_{bb}$ as is clear from the 7th row plotted in
Fig.~\ref{fig:fish_smallmap}.

A plot of $F_{bb}$ ratios (similar to the $W_b$ ratio plot of
Fig.~\ref{fig:weight}) shows that the cost of this weight redistribution
within a Fisher matrix row is a decrease in the diagonal Fisher elements
(in the $\ell=250$ to $\ell=600$ range) to 80-85\% of the exact ones.  
Not only is $F_{bb}$ suppressed then, but the larger off--diagonal
elements also lead to larger diagonal elements of $F^{-1}$.  With broader
bands (such as those used for Fig.~\ref{fig:sigmab}), the error--bar
increase due to degraded spectral resolution is not as severe.  The effect
of the larger off--diagonal elements propagates from band--to--band and is
least significant at the lower bands which are benefiting from the full
spectral resolution of the coarse submap.

Also in Fig.~\ref{fig:weight} one can see that the pixelization effects
can be fairly severe.  This is unfortunate since we only treat the
pixelization influence on the signal--correlation matrix, $S$,
approximately. Our treatment is that provided with the HEALPix package,
which assumes that the correlation between two pixels only depends on the
angular distance between them and not on their orientation.  This is an
approximation for two reasons:  the pixels are anisotropic, and their
shapes depend on their location.  The validity of the approximate
window--function can vary from submap to submap if the submaps are not
large enough to have a representative sampling of all pixel shapes. This
is another reason to use large submaps.  We take each cross--level pixel
window function to be the geometric mean of the two auto--level pixel
window functions.

Because our treatment of pixelization effects is approximate, we throw out
information from coarse submaps at a conservatively low $\ell$ value.  In
simulation $A$, for example, powers from the coarse resolution submap were
only considered for $\ell < 225$.  In the final combined results, the
higher bands only use information from the four fine resolution submaps.  
We eliminate the influence of the coarse submap on the higher bands by
inserting very large numbers into diagonal elements of the $({\cal
F}^{-1})_{\alpha l,\alpha' l'}$ matrix.  This marginalization technique is
described in Appendix A of \cite{bjk98} and can be understood as
artificially adding some noise to these particular bands so as to give
them very low weight.

The upturn in Figure ~\ref{fig:sigmab} after $\ell = 225$ where the coarse
submap information is no longer used indicates that there may be an
advantage to keeping the coarse submap information to yet higher $\ell$.  
This would require a more accurate treatment of the pixel effect on the
signal correlation function and its derivatives with respect to $C_l$.  
One way to do this, which would be fairly easy to implement and not cause
significant speed reduction, would be to avoid using pixel window
functions by calculating coarsened signal matrices directly from finer
ones.  For example, if the fine signal matrix is $s$ then the
next--coarser signal matrix, $S$, must be \be S= { {\alpha s
\alpha^T}\over 16} \ee where $\alpha_{ci}$ is one for all fine pixels $i$
in coarse pixel $c$ and zero otherwise.  Once again, we are summing four
pixels into one. The only approximations here come from approximations
made in calculating $s$.  If these approximations were acceptable for the
finer level, they will certainly be adequate for the coarser level.
Keeping the coarse level information out to higher bands may be very
important for extension to megapixel maps because it is the only other way
to improve spectral resolution besides increasing $n$ for the
highest--resolution submaps.

\section{Analysis of General Megapixel Maps}

The map from simulation $A$ has homogeneous white noise.  Below we will
discuss results from HD analysis of the map from simulation B in which the
noise is inhomogeneous but still uncorrelated.  Yet we believe HD will
work well on realistic maps with correlated noise.  In this section we
briefly make the case for the success of HD in the presence of correlated
noise and then move on to discuss how HD can be made to work for primary
maps with 100 to 1000 times more pixels than the simulation $A$ map.  We
will see that further approximations are necessary, but that they are
likely to work well.

Even though our applications of HD have only been on simulated maps with
uncorrelated noise, we believe that HD will work well on realistic maps
with correlated noise.  This is easiest to see for correlations on length
scales smaller than the size of the smallest submaps.  Longer--range noise
correlations will not be treated accurately in the analysis of the
smallest submaps.  But this does not matter because the effect will only
be on lower--$\ell$ bands where the smallest submaps do not have much
weight. The affected bands will be those determined by coarser and larger
submaps that will once again be large compared to the correlation length.
Thus the prospects for HD on maps with correlated noise are quite good.

Applying HD as we have described it to megapixel maps is prohibitively
expensive in terms of the demand on computing resources. A rough scaling
argument is sufficient to demonstrate this point. In the megapixel regime,
we are strongly dominated by the calculation of all the elements of ${\cal
F}^{-1}_{\alpha l,\alpha' l'}$. The number of elements in this matrix is
$\sim (N/n)^2 N_b^2$. On an SGI Origin 2000, the calculation of a single
element of ${\cal F}^{-1}_{\alpha l,\alpha' l'}$ takes 188 sec
$(n/2500)^3$ on a single MIPS R12000 300 MHz processor where $n$ is the
number of pixels in a submap.  Thus the wall--clock time is
\be 
\label{eqn:time1}
t \sim 1\ {\rm year} \left({500 \over n_{\rm
proc}}\right) \left({N\over 3 \times 10^6}\right)^2\left({n\over
2500}\right) \left({N_b^2\over 1000}\right)
\ee
where we have assumed the efficient use of $n_{\rm proc}$ processors
\cite{endnote4}.  Thus the need to avoid exact calculation of every
element of ${\cal F}^{-1}_{\alpha l,\alpha' l'}$ is apparent.

\begin{figure}
\plotone{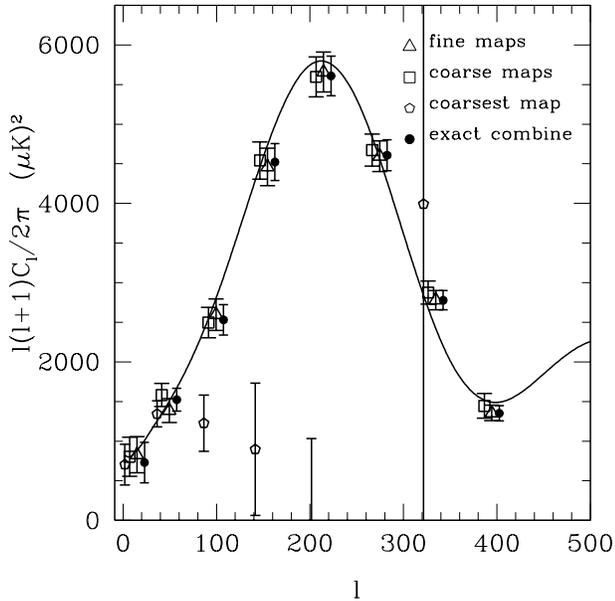}
\caption{Results from HD of the 200 by 200 pixel simulation $B$ map. There
are sixteen submaps at the finest resolution level (level 8 ($13.7'$)
pixels), four at the medium level (level 7 ($27.5'$) pixels) and one at
the coarsest level (level 6 ($55'$ pixels)). Triangles and squares
represent the result of doing a naively weighted average of the power
spectrum estimates, \ie neglecting correlations, from the sixteen fine
submaps and the four coarse submaps, respectively. Pentagons represent
results for the one coarsest submap.  Filled circles show the results of
the optimal combination of all submaps in which all power--spectrum
correlations are computed exactly. As in Fig.~\ref{fig:pow_smallmap},
points are shifted horizontally for clarity.\label{fig:pow_bigmap}}
\end{figure}

To make the case for the likely success of fast approximations to ${\cal
F}^{-1}_{\alpha l,\alpha' l'}$ we turn to the results from simulation $B$.
In Fig.~\ref{fig:pow_bigmap}, we plot four power spectra: one is the
result of optimally combining the individual power spectra; one is the
power spectrum of the coarsest submap; the other two are the result of a
{\it simple} averaging of the power spectra for the submaps within a given
resolution level as if they were independent.  Again, the solid line
represents the original input power spectrum.

We find the {\it very} good agreement between simple averaging and the
exact combination (for the highest bands) to be very encouraging because
it is strong evidence that signal correlations between non--overlapping
submaps are not very important. We certainly see they are not important in
the highest bands which are influenced only by submaps with no spatial
overlap (since the submaps are all at the same resolution level).  If any
given band is only influenced by at most two or three levels and we only
need to calculate correlations for non--zero submaps then the vast
majority of submap pairs can be ignored.  Even if some cannot be ignored,
their relative insignificance means that there are probably crude
approximations to them that will work well.

\noindent 

Calculating only the correlations between overlapping submaps at adjacent
resolution levels takes time
\be 
\label{eqn:time2}
t = 78\ {\rm hours} \left({200 \over n_{\rm
proc}}\right) \left({N\over 10^7}\right)\left({n\over
5000}\right)^2 \left({N_b\over 40}\right)\left({\Delta N_b \over
3}\right) 
\ee 

where for each of the $N_b$ bands only the nearest $\Delta N_b$ bands are
considered \cite{endnote3}.  Calculating correlations between overlapping
submap pairs whose resolution levels differ by 2 will, at most, double the
time.  Correlations between non--overlapping map pairs may be significant
but can probably be treated approximately in an insignificant amount of
time.  Development and study of these approximations is probably necessary
for practical application of HD to megapixel maps.

We also see from Fig.~\ref{fig:pow_bigmap} that even when there is a mix
of resolution levels influencing a band, using just one of those levels
provides a rough approximation. A fairly good ``quick--and--dirty''
power--spectrum estimator is the coarsest submap's power spectrum for band
1, the coarse submaps' power spectrum for bands 2 and 3, and the finest
submaps' power spectrum for bands 4 to 8.  Such an estimator has its
applications, for example, finding a $C_l$ that is close enough to optimal
that one only needs a single iteration of the exact HD procedure.

The scaling of $t$ with $N$ in Eq.~\ref{eqn:time2} is linear if $n$ is
fixed.  But if we fix spectral resolution and the area of the primary map,
then $n \propto N$ and therefore $t \propto N^3$ once again!  Or, at fixed
$N$ and primary map area, $t \propto (1/\delta l)^4$.  Our fiducial choice
above of $n=5000$ corresponds for {\it Planck} with $N=10^7$ and $r_p =
3.5'$ to $\delta_l=45$.  This may be sufficient since physical models have
fairly smooth power--spectra.  We see that the degree to which degraded
spectral resolution affects our ability to discriminate between different
models is a crucial issue for the applicability of HD to {\it Planck}.  
We remind the reader that spectral resolution is the only thing that is
significantly compromised with HD;  Fig.~\ref{fig:weight} shows the total
weight from each band is within a few percent of optimal.

\section{Comparison with other methods}

The HD method has many advantages over other fast, approximate methods.  
Perhaps the {\it chief} advantage is its ability to handle maps with
correlated noise.  Its main disadvantage is spectral resolution.  To
understand better these competitive advantages/disadvantages it is worth
spending some time discussing these other methods---especially since we
will see they are somewhat complementary and hence a hybrid approach may
be useful.

This discussion of other methods is facilitated by writing down the
following generalization of Eq.~\ref{eqn:quadest}:
\be
\label{eqn:quadestW}
C_l = \sum_{l'} {1\over 2}F^{-1}_{ll'}{\rm Tr}\left[W\left(\Delta
\Delta^{\rm T} - N\right)W{\partial C \over \partial C_{l'}}
\right]
\ee
and Eq.~\ref{eqn:fish}:
\be
\label{eqn:fishW}
F_{ll'} = {1\over 2}{\rm Tr}\left[W{\partial C \over \partial C_l}
W {\partial C \over \partial C_{l'}}\right].
\ee
These equations specify a general unbiased quadratic estimator, with pixel
pair--weighting determined by $W$.  The $F_{ll'}$ matrix is derived by
demanding that the estimator be unbiased ($\langle C_l^{\rm estimate}
\rangle = C_l$).  In general, its inverse is not equal to $\langle \delta
C_l \delta C_{l'} \rangle$ which is instead given by
\be
\label{eqn:realfishW}
{\cal F}^{-1}_{ll'} = {1\over 2} F^{-1}_{ll''}F^{-1}_{l'l'''} 
{\rm Tr}\left[A_{l''}CA_{l'''}C\right]
\ee
where $A_l \equiv W{\partial C \over \partial C_l }W$, similar to
Eq.~\ref{eqn:defA}.

For the minimum--variance estimator, $W = C^{-1}$.  The
``correlation--function'' approach (CF) of \cite{szapudi00} uses the
simpler $W=I$ in pixel space \cite{endnote1}.  Spherical--harmonic
transforming the map and averaging $|a_{lm}|^2$'s over $m$ uses $W=I$ in
spherical--harmonic space.  The multi--scale method we have just described
above likewise corresponds to a choice of $W$, although this $W$ is not
easily written down.

It is worth pointing out that the estimator for CF requires on the order
of $N_b^2N^2$ operations where $N_b$ is the number of $\ell$-bands.  One
can get rid of the $N_b^2$ factor by rewriting it as an estimator for
$C(\theta)$ in fine bins of $\theta$ and then Legendre--transforming the
result, as was done in \cite{szapudi00}.  Further computational
accelerations are possible by use of KD--tree search techniques which use
coarse--graining at large distances \cite{colombi01,moore01}.  In
addition, fast spherical harmonic transforms lead to great time--savings
in harmonic methods.

However, the simplicity of these other choices for $W$ does have
drawbacks.  Specifically, high--noise areas and low--noise areas make
equal contribution to the estimator.  To date, the success of these
methods has only been demonstrated on simulations with homogeneous white
noise. The first obvious improvement to CF is to replace
$W_{ij}=\delta_{ij}$ with $W_{ij} = 1/\sigma_i^2\delta_{ij}$ (in pixel
space) as suggested in \cite{szapudi00}.

What is less obvious is how to weight pixel pairs in the presence of
correlated noise.  This is where further development of the CF approach is
most needed. One possible route to pursue is band--diagonal choices of
$W_{ij}$ which capture the spatially--local noise correlations.
Computation with band--diagonal $W$'s can still be quite fast; they are
still order of $N^2$ as long as the bandwidth is less than $\sqrt{N}$.  
Perhaps longer--range correlations could be included in some hybrid scheme
of HD and CF.  Here CF (with band--diagonal $W_{ij}$) would be used on the
primary map and then HD would be used to calculate lower--$\ell$ values
which may have been affected by long--range noise correlations.  This
hybrid scheme also has the advantage of complementing HD where its
spectral resolution is lowest \cite{endnote2}.

Although the calculation of $C_l$ is fast with simple choices for $W$, the
calculation of the error covariance matrix (Eq.~\ref{eqn:realfishW}) is
slow; i.e. the number of operations scales with $N^3$ because of the
matrix multiplications.  One option is to estimate the errors by
Monte--Carlo methods \cite{szapudi00}.  Another is to combine the CF and
HD approaches in yet another way:  use CF as a means to produce an input
power spectrum sufficiently close to the optimal one that only a single
iteration of HD is required.

\section{Conclusions}

We have concentrated on developing a fast and reliable method for
calculating power spectra and their uncertainties from maps with $N =
10^4$ to $10^5$ pixels.  Methods that work in this regime are of immediate
practical importance.  Our tests show very good agreement with exact
methods at the lower end of our $N$ range where the exact analysis is
feasible on a supercomputer.  The HD method is the only existing method
for calculating a power spectrum and its uncertainties from general,
inhomogeneous correlated noise patterns with maps of this size in
reasonable amounts of time \cite{hivon01}.

We have not tested our method on maps with correlated noise.  But since
noise--correlations are taken into account exactly within each submap, we
expect our method to handle correlated noise effectively, unlike the other
fast methods mentioned above.  These expectations will be put to the test
soon as HD is applied to existing datasets from LDB flights, such as {\it
Archeops\/}\footnote{\texttt{http://www-crtbt.polycnrs-gre.fr/archeops/}}
and {\it TopHat\/}\footnote{\texttt{http://topweb.gsfc.nasa.gov/}}.

The local nature of the method has some advantages for controlling
contamination of the final power spectrum result.  In the extreme, one can
simply cull submaps with the largest foreground contamination.  Less
drastically, one could down--weight the power spectrum determinations from
submaps according to the suspected level of contamination.

To summarize, we have developed and investigated an HD method of
power--spectrum estimation.  We have demonstrated that for LDB--size maps
HD is sufficiently fast and insignificantly sub--optimal.  Its main
advantages over other fast methods are its generality (including its
ability to handle correlated noise) and the fact that the power spectrum
uncertainties are calculated directly.  Application to larger maps will
rely on further approximations which we expect to work well but require
further investigation.  The main disadvantage to HD is the degraded
spectral resolution at the smallest angular scales.  The impact of this
degradation on parameter--determination also warrants further
investigation.  The combination of HD with other methods may be fruitful.

{\it Madcumba}, a Fortran 90 implementation of the HD procedure, will be
made available for public use. Comments and questions should be directed
to O. Dor\'e at {\texttt dore@iap.fr}.

\acknowledgements 
O.D. is grateful to the UC Davis Cosmology group for a warm hospitality.
LK is grateful to IAP for the same.  We benefited from conversations with
J. R. Bond, J. Borrill, F. Bouchet, A. Jaffe, R. Stompor, P. Koev, D.
Vibert and R. Teyssier and the computer resources of S. Colombi and NERSC.


\begin{thebibliography}{ucsc}

\bibitem{jaffe00} E.g., A. Jaffe et al., Phys. Rev. Lett. {\bf 86}, 3475-3479
(2000). 

\bibitem{forecast} L. Knox, 
Phys. Rev. D48, 3502 (1995); 
G. Jungman, M. Kamionkowski, A. Kosowsky, and D. Spergel 1996, Phys.
Rev. D {\bf D54}, 1332 (1996); J. R. Bond, G. Efstathiou, and M. Tegmark
Mon. Not. Roy. Astron. Soc., {\bf 291}, L33 (1997);
D. Eisenstein, W. Hu
and M. Tegmark, 
Astrophys. J. {\bf 504}, 57L (1998).

\bibitem{bcjk} J.~R. Bond, R. Crittenden, A.~H. Jaffe and L. Knox, 
Computing in Science and Engineering, vol. 1, no. 2, 21 (1999).

\bibitem{borrill99} J. Borrill, Phys. Rev. D {\bf 59}, 027302 (1999).

\bibitem{dore01} O. Dor\'e, R. Teyssier, F.R. Bouchet, D. Vibert,  
astro-ph/0101112, see also http://ulysse.iap.fr/download/mapcumba

\bibitem{wandelt} B.D. Wandelt, E. Hivon \& K. G{\'o}rski,
astro-ph/0008111;
astro-ph/9808292

\bibitem{szapudi00} I. Szapudi, S. Prunet, D. Pogosyan, A. Szalay and J.R. Bond, 
astro-ph/0010256

\bibitem{oh99} S.P. Oh, D.N. Spergel and G. Hinshaw,
Astrophys. J. {\bf 510}, 551 (1999).

\bibitem{wandelt01} B. Wandelt 2001, Proceedings of
MPA/MPE/ESO Conference "Mining the Sky", July 31 - August 4, 2000,
Garching, Germany, astro-ph/0012333, astro-ph/0012416

\bibitem{bjk98} J.R. Bond, A.H. Jaffe \& L. Knox, Phys. Rev. D {\bf 57},
2117 (1998).

\bibitem{tegmark97} These equations were independently derived as
the optimal, unbiased quadratic estimator in 
M. Tegmark, Phys. Rev. D {\bf 55}, 5895 (1997).

\bibitem{endnote5} This can also be understood as the usual
problem of localizing simultaneously in position and momentum:
M. Tegmark,  Mon.Not.Roy.Astron.Soc. {\bf 280}, 299-308 (1996).

\bibitem{borrill01} J. Borrill and P. Koev 2001, in preparation.

\bibitem{GoHi98} G{\'o}rski E.K., Hivon E., Wandelt B.D.
in proceedings of the MPA/ESO Garching Conference 1998, eds Banday
A.J., Sheth K. and L. Da Costa and
http://www.eso.org/~kgorski/healpix/

\bibitem{jaffe01} A. Jaffe, private communication.

\bibitem{bjk00} J.R. Bond, A.H. Jaffe \& L. Knox, \apj 
{\bf 533}, 19 (2000).

\bibitem{seljak96} U. Seljak \& M. Zaldarriaga, \apj, {\bf 469}, 437, 1996

\bibitem{endnote4}We
tested the scaling with $n_{\rm proc}$ by running {\it Madcumba} using up
to
$77$ processors.  

\bibitem{endnote3}We remind the reader that application of HD to
megapixel and larger maps requires some way to calculate $N$ for the
sub--maps from the time--ordered data.  This could be accomplished
by the method we briefly described in subsection IID, which will
be described in more detail in \cite{borrill01}.

\bibitem{endnote1} That
the correlation--function approach can be regarded as a quadratic
estimator with sub--optimal weighting was emphasized in \cite{szapudi00}.

\bibitem{colombi01} S. Colombi {\it et al.} 2001, in preparation.

\bibitem{moore01} A. Moore {\it et al.} 2001, Fast Algorithms and
Efficient Statistics: N-point Correlation Functions, Proceedings of
MPA/MPE/ESO Conference "Mining the Sky", July 31 - August 4, 2000,
Garching, Germany, astro-ph/0012333

\bibitem{endnote2} The speed--up with KD--tree search techniques
will also lead to some spectral resolution degradation.

\bibitem{hivon01} A possible exception is 
a monte--Carlo pseudo-$C_l$ method to be described in
E. Hivon {\it et al.} 2001, in preparation.

\end{thebibliography}
\end{document}